# Electronic properties of topological insulator candidate CaAgAs

Jayita Nayak[1‡*], Nitesh Kumar[1‡], Shu-Chun Wu[1], Chandra Shekhar[1], Jörg Fink[1,2], Emile D. L. Rienks[2,3], Gerhard H. Fecher[1], Yan Sun[1] and Claudia Felser[1*]

[1] Max Planck Institute for Chemical Physics of Solids, Nöthnitzer Str. 40, D-01187 Dresden, Germany.

[2] Leibniz Institut für Festkörper- und Werkstoffforschung Dresden, Helmholtzstrasse 20, D-01171 Dresden, Germany.

[3] Institute of Solid State Physics, Dresden University of Technology, Zellescher Weg 16, 01062 Dresden, Germany.

[‡] Equal contribution
* Claudia.Felser@cpfs.mpg.de, Jayita.Nayak@cpfs.mpg.de

## Abstract

The topological phases of matter provide the opportunity to observe many exotic properties, such as the existence of two-dimensional topological surface states in the form of Dirac cones in topological insulators and chiral transport through the open Fermi arc in Weyl semimetals. However, these properties affect the transport characteristics and, therefore, may be useful for applications only if the topological phenomena occur near the Fermi level. CaAgAs is a promising candidate for which the *ab-initio* calculations predict line-nodes at the Fermi energy. However, the compound transforms into a topological insulator on considering spin-orbit interaction. In this study, we investigated the electronic structure of CaAgAs with angle-resolved photoelectron spectroscopy (ARPES), *ab-initio* calculations, and transport measurements. The results from ARPES show that the bulk valence band crosses the Fermi energy at the Γ-point. The measured band dispersion matches the *ab-initio* calculations closely when shifting the Fermi energy in the calculations by −0.5 eV. The ARPES results are in good agreement with transport measurements, which show abundant *p*-type carriers.

The experimental realization of the topological insulating (TI) phase led to the explosion of current research activities [1, 2]. The search for new topological states of matter has emerged as an active research field in condensed matter physics. After the discovery of topological insulators, researchers proposed and experimentally investigated many new semimetallic or metallic phases, which host non-trivial band topology with vanishing densities of states near the Fermi energy [3-10]. In general, the topological semimetals are distinguished in two major classes depending on whether conduction and valence bands coincide at a point or along a line in the three-dimensional (3D) momentum space. This results in Dirac and Weyl semimetals, or nodal line semimetals, respectively. Band touching points in the topological semimetals are protected by certain symmetries, analogous to TIs, and thus are robust against external perturbations. 3D Dirac semimetals are experimentally realized, for example, in the compounds $Na_3Bi$ [11, 12] and $Cd_2As_3$ [9, 13]. The violation of either inversion or time-reversal symmetry in Dirac semimetals leads to the occurrence of Weyl semimetals. These were verified experimentally, for example, in transition metal monophosphides [8, 14, 15]. Quite recently, many topological nodal line semimetals have been predicted but only a few of them are experimentally confirmed by ARPES measurements [10, 16]. *Ab-initio* calculations reveal that the nodal points are quite fragile against spin-orbit interaction [17, 18] which will drive the system into a topological insulator or a Weyl semimetal phase by opening up of a gap at the nodal points [19]. Due to strong spin-orbit interaction rare earth based pnictide LaBi is a TI which is verified experimentally [20].

First-principles calculations by Yamakage *et al.* revealed that the non-centrosymmetric compound CaAgAs might host a topological semimetallic state. A nodal line appears at the Fermi energy in the absence of SOC [21]. On incorporating SOC, it transforms into a TI where the Fermi energy resides within the inverted band gap. Quite recently, Wang *et al.* claimed the presence of a non-$k_z$ dispersive surface band that coincides with the topological surface state in CaAgAs [22]. On the other hand, Takane *et al.* concluded from bulk sensitive soft X-ray photoemission that CaAgAs belongs to the family of nodal line semimetals [23]. We study the electronic structure of CaAgAs by angle-resolved photoelectron spectroscopy (ARPES), *ab-initio* calculations, and transport measurements, and establish that the states at the Fermi energy are dominated by the bulk valence band because of self-doping in the single crystals.

CaAgAs single crystals were grown by the flux method with bismuth as an external flux. Ca, Ag, As, and Bi pieces were weighed according to the composition $(CaAgAs)_{0.1}Bi_{0.9}$ and transferred into an alumina crucible. It was then vacuum-sealed inside a quartz tube. The content was heated to 1100 °C and kept at this temperature for 24 h. Then, it was slowly cooled at 1.5 °C/h to 700 °C. At this temperature, the flux was decanted off to obtain single crystals of CaAgAs. The quality of the crystals was confirmed by Laue X-ray diffraction (XRD), and the composition was determined by energy-dispersive analysis of X-rays (EDAX). The ARPES measurements were carried out at the UE112-PGM2b beamline of BESSY (Berlin) using the $1^3$-ARPES end station, which is equipped with a Scienta R4000 energy analyser. All measurements were performed at a temperature of 1 K at photon energies ranging from 55 to 150 eV. The sample was cleaved *in situ* at a low temperature (~30 K). The total energy resolution was approximately 4 meV, and the angular resolution was 0.2°.

The band structure was calculated by the Vienna *ab initio* simulation package (VASP) using projected augmented wave (PAW) potentials [24, 25]. The generalized gradient approximation (GGA) [26] was used for the exchange-correlation functional. The surface electronic states are investigated by a tight-binding model, which was constructed from maximally localized Wannier functions [27] for Ag-*s*,*d*, As-*p*, and Ca-*d* orbitals.

The non-centrosymmetric hexagonal pnictide CaAgAs crystallises in a ZrNiAl-type structure. The crystal structure is presented in Fig. 1(a); it belongs to the space group $P\bar{6}2m$ (No. 189). The important aspect of this space group is that it contains mirror reflection symmetry but has no centre of inversion. The X-ray Laue diffraction of a CaAgAs single crystal is shown in Fig. 1(b). The Laue pattern is overlaid with the simulated pattern (red dots) according to the space group $P\bar{6}2m$ in Fig 1(c). The bulk Brillouin zone and the high-symmetry directions are shown in Fig. 1(d). The two-dimensional projection of the (0001) surface Brillouin zone is also included in Fig. 1(d). The bulk band structure of CaAgAs was calculated along the path Γ-M-K-Γ-A-L-H-A. The result without considering SOC is shown in Fig. 1(e). The calculated band structure points on the appearance of a nodal line semimetallic phase in the absence of SOC. It was found that the line node is centred on Γ for the directions ΓM (Σ) and ΓK (Λ). The $p_z$ orbital at the As atoms contribute mainly to the conduction band near Γ, whereas the valence band is formed from the $p_x$, $p_y$ orbitals of As along with the *s* orbital of Ag. Thus, a band inversion occurs between As $p_x$, $p_y$, and $p_z$ orbitals. Further, SOC was used in the calculations to explain its influence on the nodal lines.

As seen from Fig. 1(f), now the line nodes become gapped along the ΓM and ΓK directions and drive the system to the TI phase. Wannier functions have been used to identify the topological order [28, 29]. The result indicates a non-trivial topology, which is consistent with previous work [21]. Further, we confirmed the TI phase by direct surface-state calculations with Green's function on the (0001) surface of CaAgAs with $Ca_3As$ termination. The Dirac point is 0.1 eV below the charge neutral point; thus it is still inside the inverted band gap.

We performed ARPES measurements on the CaAgAs (0001) surface. The state of the surface was characterised by core-level spectroscopy (Fig. 2(a)). The Ca 3*p* semi-core level is identified at 24-eV binding energy; its spin-orbit splitting is not resolved. The maximum of the As 3*d* peak appears at 39.48 eV with a spin-orbit splitting of 0.6 eV. The intensity of the Ag 4*p* peak (shown in the inset of Fig. 2(a)) is strongly suppressed. The absence of any impurity peak and the appearance of sharp Ca and As peaks in the core-level spectra hint on the high quality of the crystal used for spectroscopic investigations. Fig. 2(b) shows an iso-energy surface of CaAgAs measured with a photon energy of 85 eV. The Fermi surface appears as a circular hole pocket centred at $\bar{\Gamma}$ with $k_F$=0.2±0.02. The red dotted box indicates the hexagonal surface Brillouin zone of CaAgAs marked with the high symmetry points $\bar{\Gamma}, \bar{M}$ and $\bar{K}$. A constant-energy surface at −0.35 eV is shown in Fig. 2(c) and reveals that the radius of the hole pocket increases (marked by the black circle for the guidance of eye) with increasing binding energy and merges with those around the neighbouring $\bar{\Gamma}$ points. The calculated Fermi surface reproduces the experimental data after an energy shift of −0.5 eV away from the charge neutral point (Fig. 2(d), (e)). The calculated wave vector of $k_F$=0.19 Å$^{-1}$ nicely matches with the ARPES results. The shift is probably caused by a small self-doping. To estimate the amount of self-doping required for shifting the chemical potential by -0.5 eV we have theoretically evaluated from the integral of the density of states. We found that 0.105 electron needs to be removed per unit cell to shift the Fermi energy by -0.5 eV. This may be achieved either by creating 1% Ag vacancies corresponding to the composition $CaAg_{0.99}As$ [22] or may be attained by Ca or As vacancies corresponding to the compositions $Ca_{0.95}AgAs$ or $CaAgAs_{0.98}$. Indeed, other variations of the composition that result in electron depletion are also possible. Both experiment and calculations demonstrate that the Dirac point is well above $E_F$. Further, the crystal exhibits predominantly *p*-type behaviour, which is consistent with the transport measurements that will be discussed later. It should be noted that it was not possible to shift the Dirac point below the Fermi energy by Cs deposition.

We measured the band dispersion along the path $\overline{M} - \overline{\Gamma} - \overline{M}$ (shown in Fig. 2(f)). The ARPES data were converted to the second derivative (Fig. 2(g)) to identify the bands clearly and indicate the appearance of another hole-like band below 0.2 eV inside of the outer hole-like band. Interestingly, the inner band is also evident in the calculated spectra (Fig. 2(h)). In contrast to that, the constant energy surface at −0.35 eV does not resolve the inner band due to the overlap with the outer band. The band dispersion (Fig. 2(i), (j)) along $\overline{K} - \overline{\Gamma} - \overline{K}$ is also consistent with the calculation (Fig. 2(k)). In this context, another important observation is that both the inner (Fig. 2(f-g)) and outer (Fig. 2 (i-j)) bands are asymmetric and the asymmetry is related to the matric element effect in the photoemission, that is, the intensity will depend on the angle between momentum of the outgoing electron and the electric field vector of the photons.

We performed photon-energy-dependent ARPES measurements in a wide photon energy range between 60 to 87 eV along $\overline{M} - \overline{\Gamma} - \overline{M}$ direction as illustrated in Fig. 3(a)–(d). We find that both the inner and outer hole pockets centred at $\overline{\Gamma}$ change significantly with the change of photon energy, which indicates that the hole pockets result from bulk bands. The observed strong dispersion with photon energy is expected for bulk bands, whereas surface states should not disperse with photon energy because they do not disperse with $k_z$. We have compared our ARPES data (Fig. 3(e)) at 90 eV directly with the calculated band dispersion (Fig. 3(f)) along M-Γ-M direction, which is in fair agreement since the calculated spectra attribute to the balk bands. On the other hand, Wang *et al.* claimed the presence of topological surface states along the edge of the bulk hole pockets. To settle the issue of whether there is any surface state, we plotted the intensity $I(k_x, k_z)$ distribution (Fig. 2(g)) at the Fermi energy and found a strong $k_z$ dependence. This demonstrates that the observed states are bulk states, in stark contrast to the previous report. We have considered inner potential $V_0$ to be 13 eV for the $k_z$ scaling. In addition, the calculation rejects the possibility of any surface state at the edge of the bulk hole bands and confirms that the topological surface states are strictly restricted between the so-called "nodal points" (Fig. 2(h), (k)). Here, it is important to mention that the calculation is in excellent agreement with the reports by Yamakage *et al* [21].

Moreover, a second feature which is observed at $k_X=0.88$ Å$^{-1}$ in Fig 2(m) represents the next $\overline{\Gamma}$ point and also exhibits a strong photon energy dependence. Seemingly, the feature at the second $\overline{\Gamma}$ point reveals a $k_z$ offset of ~0.35 Å$^{-1}$. This offset is predominantly caused by the difference in the matrix element of bulk bands appearing at the two $\bar{\Gamma}$ points as evident in the photon energy dependent ARPES spectra.

The measured transport properties of CaAgAs are in good agreement with the ARPES and the earlier transport data available [30]. The electrical resistivity at 2 K is $7.4\times10^{-5}$ $\Omega$cm. It decreases with decreasing temperature and saturates at a low temperature, typical of a metal (Fig. 4(a)). This is in contrast to the topological insulators, where the Fermi energy should be inside the inverted band gap and the resistivity should increase with decreasing temperature with a plateau at a low temperature due to the topological surface states. The Hall resistivity $\rho_{yx}(B)$ demonstrates the existence of hole-type charge carriers from its positive slope (see Fig. 4(b)). This is expected when the valence band crosses the Fermi energy as verified previously by ARPES. The behaviour is highly robust against temperature; there is only a small change up to 300 K. We extracted the hole carrier concentration, $n$ and Hall mobility, $\mu$ from the slope (Hall coefficient, $R_H$) of $\rho_{yx}(B)$ and $\rho_{xx}(B=0T)$ by using the relations $\mu = R_H/\rho_{xx}$ and $n = 1/e.R_H$ at 2 K, the values are $1.67\times10^{20}$ cm$^{-3}$ and 505 cm$^2$V$^{-1}$s$^{-1}$, respectively. The value of $R_H$ is $3.75\times10^{-6}$ $\Omega$cmT$^{-1}$ at 2 K. The carrier concentration is nearly independent of temperature, indicating that the shape and the size of the Fermi surface remain intact. This carrier concentration is in fair agreement with the carrier density obtained from first principles calculations ($2\times10^{20}$ cm$^{-3}$). The theoretical carrier density was calculated after shifting the Fermi energy by -0.5 eV in order to match the ARPES data. In the framework of the Drude model, considering a spherical hole-type Fermi surface, the carrier concentration from Hall resistivity gives a Fermi wave vector of 0.17 Å$^{-1}$, which is in close agreement with that observed in ARPES (0.2±0.02 Å$^{-1}$).

In summary, we studied the electronic properties of CaAgAs by employing ARPES, *ab-initio* calculations, and transport measurements. In the present work, we verified that the bands at the Fermi energy have bulk characteristics. The agreement between ARPES and *ab-initio* calculations reveal that the Dirac point is approximately 0.4 eV above the Fermi energy. Dominating hole-type carriers from the Hall resistivity data further support our claim.

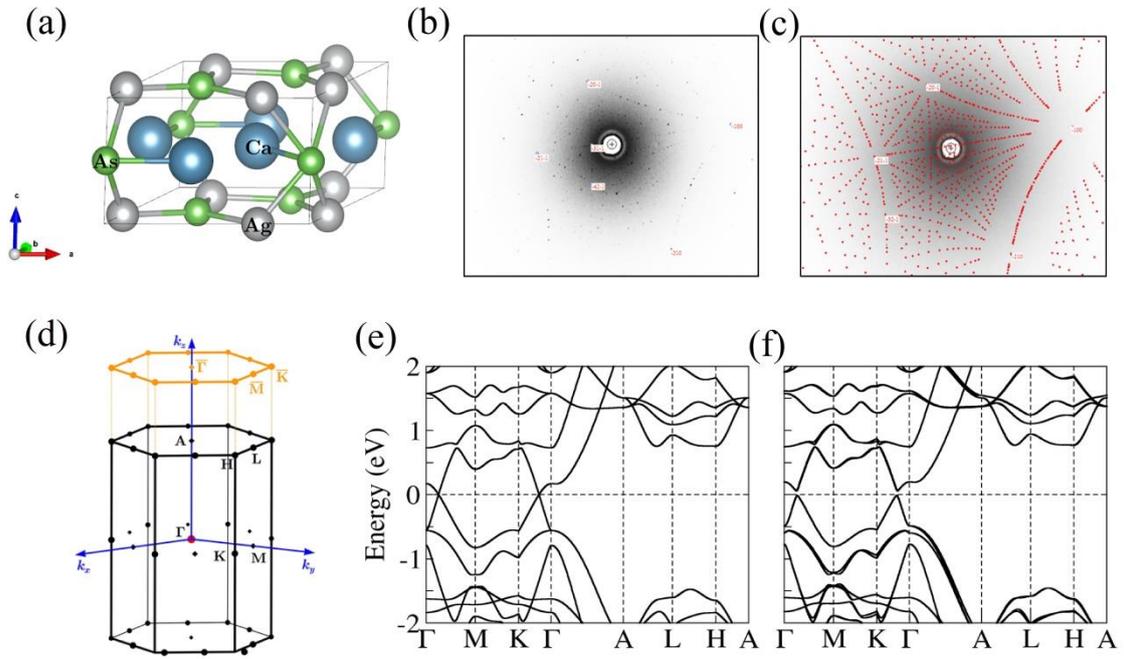

**Fig. 1: Crystal structure and band structure.** (a) Hexagonal crystal structure of CaAgAs, (b) x-ray Laue diffraction pattern of the single crystal, (c) simulated Laue diffraction pattern overlapped with the experimental data, (d) bulk and (0001) projected surface Brillouin zone. Calculated bulk band structure of CaAgAs without (e) and with (f) spin orbit coupling.

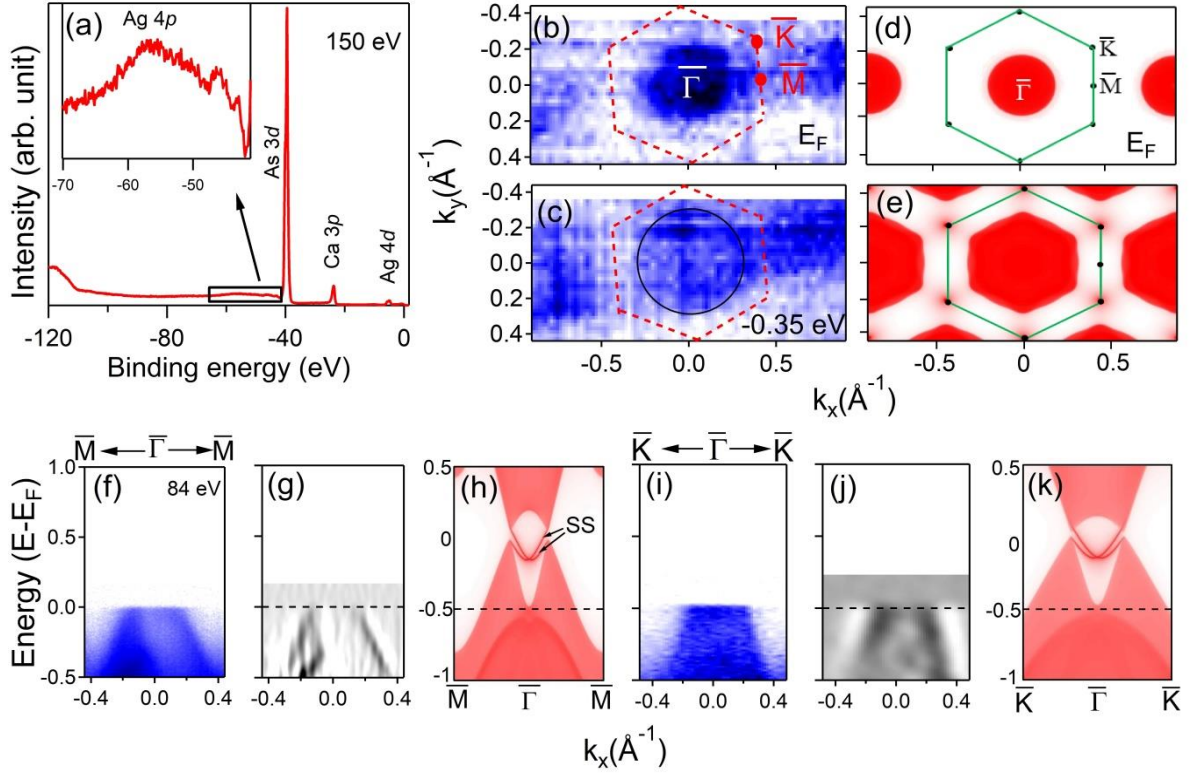

**Fig. 2: ARPES measurement of CaAgAs (0001).** (a) Semi-core-level spectrum of CaAgAs acquired with 150-eV excitation energy showing the Ca 3*p*, Ag 4*d*, and As 3*d* peaks. The inset shows the Ag 4*p* peak on an expanded scale. (b) Fermi surface of CaAgAs measured with 85-eV photon energy, marked with Brillouin zone and high symmetry points, (c) constant-energy surface of CaAgAs at −0.35-eV binding energy, and (d), (e) calculated constant-energy surface at 0 eV and −0.35 eV, respectively.  (f) ARPES spectrum, (g) second derivative, and (h) calculated spectra along $\bar{M} - \bar{\Gamma} - \bar{M}$ respectively; (i)–(k) similar results as (f)–(h) but measured along $\bar{K} - \bar{\Gamma} - \bar{K}$.

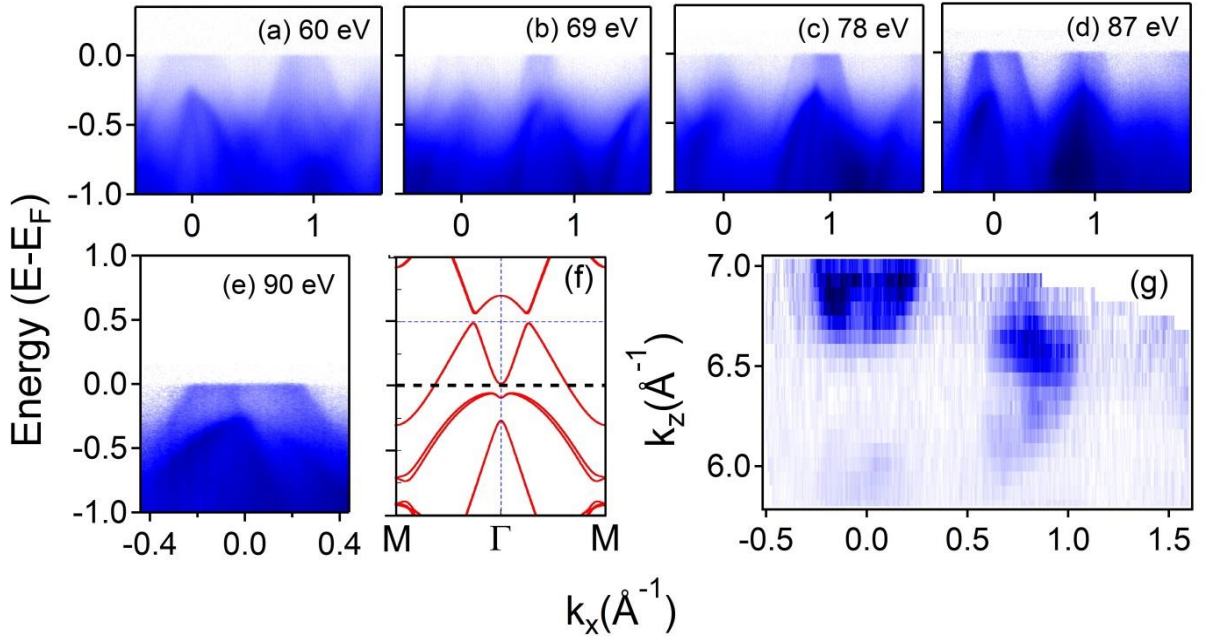

**Fig. 3: Photon-energy-dependent ARPES measurement of CaAgAs.** (a)-(d) ARPES spectra measured along $\overline{M} - \overline{\Gamma} - \overline{M}$ with photon energies from 60 eV to 87 eV with a step size of 9 eV. (e) ARPES spectrum at 90 eV compared with calculated spectrum (f) along M-Γ-M direction. (g) ARPES spectral intensity maps in the $k_x$-$k_z$ plane at the Fermi energy.

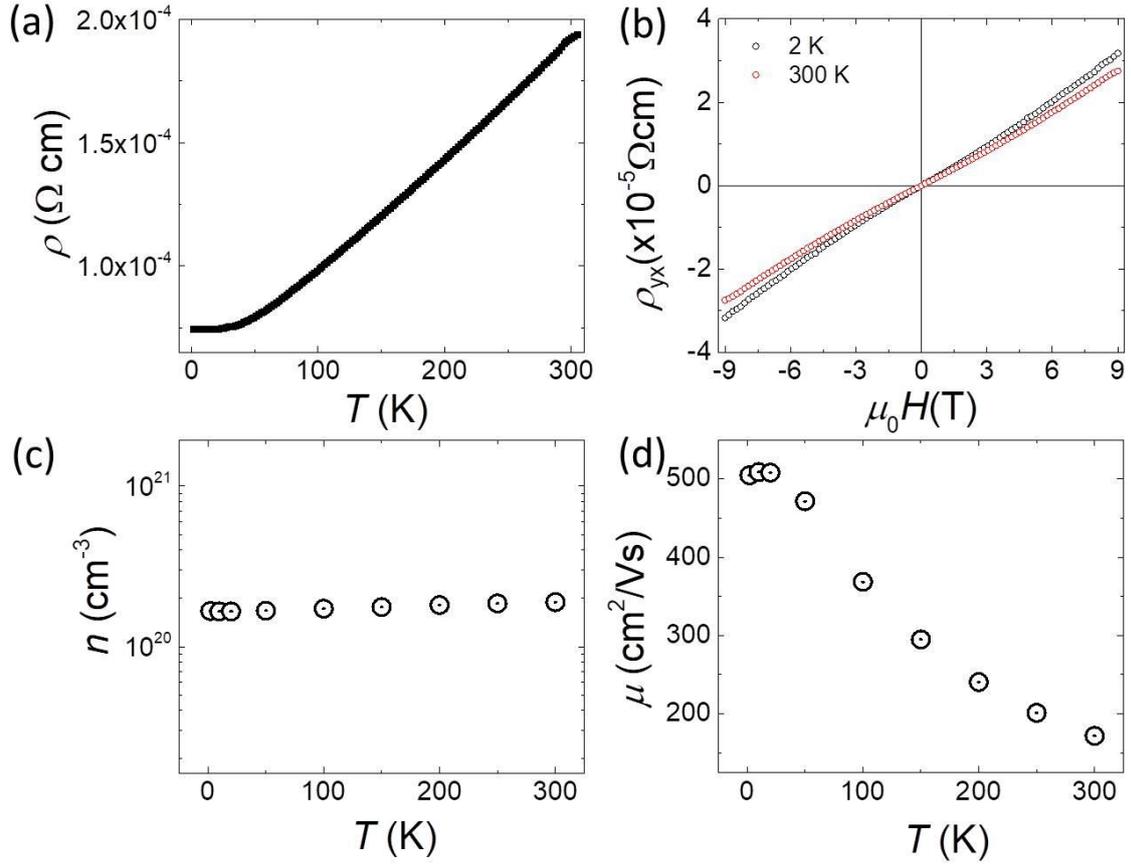

**Fig. 4: Transport properties of CaAgAs:** (a) $\rho_{xx}(T)$ data at zero magnetic field. (b) Field-dependent Hall resistivity $\rho_{yx}(B)$ at 2 K and 300 K. Temperature dependence of (c) carrier concentration and (d) mobility.

12